# Mode coupling, bi-stability, and spectral broadening in buckled carbon nanotube resonators

S. Rechnitz[1†], T. Tabachnik[1†], M. Shlafman[†], S. Shlafman[†], and Y. E. Yaish[*†]

[†]Andrew and Erna Viterbi Faculty of Electrical Engineering, Technion, Haifa, Israel.

[1]These authors contributed equally to this work.

**Abstract**

Bi-stable mechanical resonators play a significant role in various applications, such as sensors, memory elements, and quantum computing. While carbon nanotube (CNT) based resonators have been widely investigated as promising nano electro-mechanical devices, a bi-stable CNT resonator has never been demonstrated. Here, we report a new class of CNT resonators in which the nanotube is buckled upward. We show that a small upward buckling yields record electrical frequency tunability, whereas larger buckling can achieve Euler-Bernoulli (EB) bi-stability, the smallest mechanical resonator with two stable configurations to date. We believe that these recently discovered CNT devices will open new avenues for realizing nano-sensors, mechanical memory elements and parametric amplifiers. Furthermore, we present a three-dimensional theoretical analysis revealing significant nonlinear coupling between the in-plane and out-of-plane static and dynamic modes of motion, and a unique three-dimensional EB snap-through transition. We utilize this coupling to provide a conclusive explanation for the low quality factor in CNT resonators at room temperature, key in understanding dissipation mechanisms at the nano scale.

**Main text**

A bi-stable system is the underlying operation principle of modern technology. It is the basic building block for computing, memories, and digital electronics. In micro electro-mechanical systems (MEMS) technology[1], bi-stability is achieved by an arch shaped beam with two symmetric possible configurations: either buckled up (or right) or down (or left). The transition between the two buckled configurations, named Euler-Bernoulli (EB) snap-through (ST) bi-stability[2], is a fascinating example for non-linear behavior where the compression of the beam decreases its effective spring constant until it vanishes. The transition between the two configurations is controlled by an external force (usually electrostatic) and near this critical transition point, any small force perturbation results in a significant mechanical response, utilized for the realization of ultra-sensitive force, acceleration, and position sensors. Reducing the dimensions of a MEMS resonator to the nano scale improves its performance[3–6] and enables observation of quantum effects that are not accessible via MEMS devices[7].

Carbon nanotube (CNT) based resonators have been widely investigated for sensing[8,9], signal processing[10], and quantum research[11–15]. However, a bi-stable CNT resonator has never been demonstrated. In this work, we realize for the first time, the EB snap-through buckling instability in novel suspended CNT resonators and investigate, both theoretically and experimentally, the static and the dynamic behavior of the system, and visualize the complete three-dimensional CNT motion. In fact, we present the first devices in which the CNT is initially buckled upward. This new configuration enables unique out-of-plane static motion which results in completely different behavior than traditional CNT resonators. The realization of this new type of devices opens new avenues, some of which we report here:

robust bi-stability for long-term endurance of the device, 10-fold enhancement over the best electrical frequency tunability reported, and strong mode coupling explaining the anomalous dissipation of CNTs at room temperature.

An important characteristic of CNT resonators is the emergence of nonlinear behavior even at a nanometric vibrational amplitude[16,17]. In devices exhibiting EB buckling, the geometric nonlinearity[18,19] becomes more significant since we can observe the transition from compression to stretching. This transition is accompanied by large modulation of the resonance frequencies and strong nonlinear coupling between different modes via the built-in strain. We utilize this induced coupling by the axial strain between the in- and out-of-plane mechanical motions to provide decisive explanation for the long-standing problem of low quality factor of CNT resonators at room temperature, in agreement with the fluctuation broadening theory[20].

**Device fabrication and results**

Fig. 1a presents the geometric structure of our devices. The fabrication is based on the technique developed recently for self-aligned local gate suspended CNT devices[21]. Briefly, CNTs are grown by chemical vapor deposition (CVD) to create a contact between the source and drain (SD) electrodes, while suspended above a local-gate electrode (Fig. 1a, see methods section and supplementary information for fabrication details).

All measurements were conducted in a vacuum chamber under a constant pressure of $10^{-5}$ Torr at room temperature. For the resonance frequencies measurements, the actuation of the CNT is done electrostatically by the local-gate, and the detection is done by the frequency mixing technique[22], as illustrated in Fig. 1a.

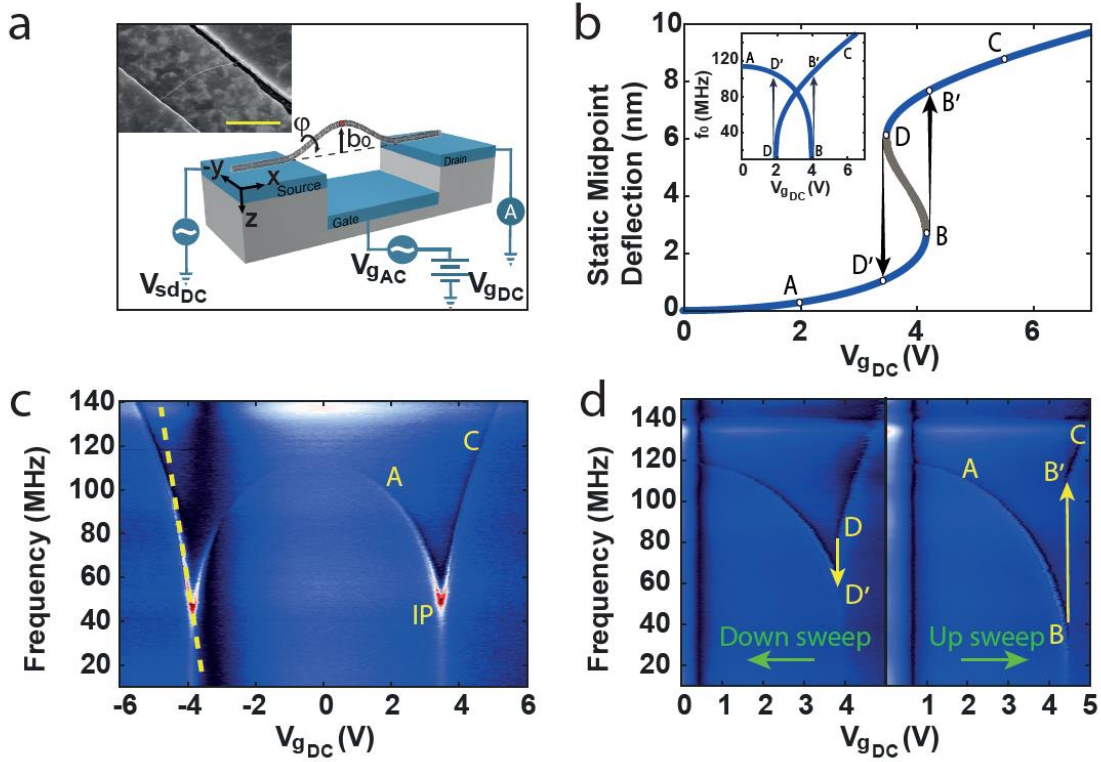

**Fig. 1: Suspended CNT based bi-stable NEMS device.** *(a) Schematic layout of the device, coordinates system and experimental setup for the resonance frequency measurement. Note that our coordinates system is such that the positive z direction points downward. Inset: SEM image of a typical device, displaying initial upward buckling. Scale bar - 1μm. (b) Theoretical 2D model for the midpoint static deflection (red dot in a) as a function of the DC gate voltage. The blue line represents a stable solution while the gray is unstable. Inset is the corresponding theoretical resonance frequency dependence on the DC gate voltage. (c) Resonance frequency measurement of device I, displaying continuous transition via an inflection point (IP), exhibiting $df_0/dV_G$ ~100 MHz/V (dashed yellow line). (d) Upward (right) and downward (left) sweeps measurement of device II, exhibiting the snap-through and release phenomena, resulting in a "jump" of nearly 80 MHz in the resonance frequency.*

Figs. 1c and 1d present resonance frequency measurements of two devices (devices I and II, respectively), revealing a qualitative different behavior from previously reported suspended CNT resonators[22,23]. The "classical" suspended CNT resonance frequency increases with the gate voltage due to increase in tension. In our devices, however, the resonance frequency exhibits a substantial decrease until a noticeable minimum. The transition from "fall" to "rise" appears either in the form of a sharp dip via an inflection point (IP, Fig. 1c) or by a large "jump" in the resonance frequency (~80MHz in Fig. 1d). The IP in Fig.1c is characterized by extremely high tunability (marked by the dashed line) with a slope of 100 MHz/V, the highest electrostatic tunability to date of any NEMS/MEMS device[22,24]. High $df_0/dV_g$ ratio is essential for achieving ultrasensitive nano-sensors[3], realizing nanomechanical computing[25], and for obtaining high-gain parametric amplification and self-oscillation.

Fig. 1b presents results from a standard 2D theoretical model that should predict the static response of a buckled beam exhibiting ST phenomenon. Initially, the CNT is buckled upward. Applying a DC gate voltage attracts the CNT towards the local-gate, generating compression and therefore softening the CNT spring constant, which causes the decrease in the resonance frequency (point A in Figs. 1b-d). If the initial buckling is above a critical instability, an unstable solution is formed (marked in gray, Fig. 1b). Then, the transition between the two stable configurations (in our case, from initial upward buckling to downward buckling) occurs through an abrupt mechanical transition, known as the snap-through Euler-Bernoulli buckling transition[19,26]. This mechanical "jump" causes a change in the spring constant, which translates into a discontinuity in the resonance frequency ($B \rightarrow B'$ in Figs. 1b,1d). After the snap, decreasing the applied voltage, a second critical point is reached (D) resulting in an abrupt transition back to the original geometric configuration, known as release ($D \to D'$). The ST and release points may occur for different static loads (as is depicted in Fig. 1b) which then translates into a hysteretic response, evidenced in both the DC conductance measurement (see supplementary text) and the resonance frequency measurement (Fig. 1d).

**Discussion**

We divide our devices into four different categories (Fig. 2 and S3). In the first category (device III, see Fig. 2a), the CNT is initially buckled downward ($b_0>0$), thus exhibiting the common behavior of CNT resonators which has been widely studied[22,24,27].

The other categories, however, deal with the case of initial upward buckling ($b_0<0$), a regime never explored before in CNT resonators. These classes are characterized by the ratio between $b_0$ and the critical initial buckling, $b_c<0$, for which the ST instability is formed.

In the second group, the initial height is still minor compared to the critical value, meaning $|b_0| \ll |b_c|$. The resonance frequency gate dependence in this regime is characterized by a mild dip in the resonance frequency measurement (device IV, Fig. 2b). As $|b_0|$ increases, we enter the third category, in which the mild dip sharpens (devices V and I, Figs. 2c and 1c, respectively), but the transition is still continuous. When $|b_0|$ increases above the critical value, we enter the fourth type ($|b_0|>|b_c|$), characterized by a hysteretic static and dynamic response. As a result, a noticeable "jump" in the resonance frequency is observed (devices II and VI in Figs. 1d and 2d, respectively).

Unlike previous CNT bi-stable devices that relied only on pull-in[4], which deteriorate after only several switches[28], our ST transition is extremely robust. The snap-through transition occurs in every DC gate sweep. Hence, during a resonance frequency measurement, it is repeated hundreds of times. For example, we extracted an average ST voltage from the measurement in Fig. 1d of $V_{ST}$=4.39V and a standard deviation of $\sqrt{\langle \Delta V_{ST}^2 \rangle}$= 0.02V. The relative deviation in the snap-through voltage ($\sqrt{\langle \Delta V_{ST}^2 \rangle}/V_{ST} = 0.46\%$) in our CNT resonators is significantly smaller than in previous mentioned bistable nanobeams ($V_{ST}$~50V, $\sqrt{\langle \Delta V_{ST}^2 \rangle}/V_{ST} = >10\%$)[29]. This shows that the ST transition in our CNT resonators is extremely sharp and requires significantly lower power consumption.

While these results are highly encouraging, they leave us with a puzzle. According to the 2D model (Fig. 1b), the ST transition occurs at extremum points, which symbolize zero spring constant and hence zero resonance frequency ($f_0$=0). However, looking closer at Figs. 1d and 2d (of devices in the fourth category), we note that the "jump" does not take place at $f$=0, but at 20-60MHz above zero. In addition, for the down sweep in Fig. 1d (left) we observe a negative jump, which cannot be explained by the naïve model. To solve this puzzle, we had to realize that the model in Fig. 1b assumes that the CNT movement is constrained to the xz-plane (in-plane), where in fact, there is a "hidden" out-of-plane component responsible for our observations.

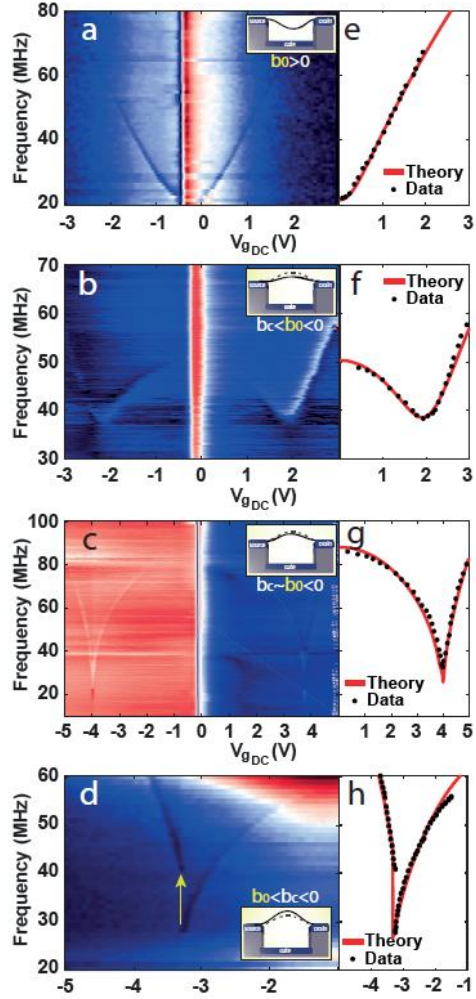

**Fig. 2: Influence of initial buckling on the dynamic response.** *(a-d) Representative resonance frequency measurements of devices III-VI for the different categories, where (a) $b_0>0$, (b) $b_c \ll b_0<0$, (c) $b_c \lesssim b_0<0$, and (d) $b_0<b_c<0$. The insets are schematics of the initial shape of the CNT (solid arch) illustrating the different classes relative to the critical value $b_c$ (dashed arch). (e-h) Modelling the dynamic response of the same devices in (a-d), respectively. The solid lines are the theoretical fitting to the experimental data (black dots). The initial CNT heights extracted from the fit are: (e) $b_0=10nm$ for device III, (f) $b_0=-6nm$ for device IV, (g) $b_0=-38nm$ for device V, and (h) $b_0=-45nm$ for device VI, in agreement with our qualitative interpretation.*

In order to understand the origin of the different classes of devices and their characteristics, we model our devices as doubly clamped beams[19,26,30]. Since the CNT cross section is circular, we must consider the out-of-plane as well as the in-plane motion[16,31]. The out-of-plane motion and the electrostatic force towards the local-gate impose torque along the CNT axis[30], forming a gate-dependent torsion. Thus, the tube develops a twist during its static motion towards the local-gate. As a result, we formulated a 3D theoretical model in which the Euler-Bernoulli equations of motion are solved considering all three degrees of freedom: in-plane (z), out-of-plane (y) and twist ($\varphi$) (see supplementary text). We should emphasize the fundamental difference whether the CNT is slacked downward (type I) or buckled upward (types III-V), since only the latter can result in static out-of-plane motion. While many CNT studies took into account out-of-plane dynamics[16], we are the first to consider out-of-plane initial and static deflection. This type of motion is new both for suspended CNTs analysis (traditionally slacked downward) as well as for buckled beams analysis, in which the y (width) and z (thickness) beam dimensions are usually different, restricting the motion to be solely in-plane (width$\gg$thickness). Our simulations yielded excellent agreement with the experimental results (Figs. 2e-h and 3c-d) and allowed us to visualize the complete and unique CNT motion (Figs. 3e-f).

Regarding the puzzle raised earlier, as to why does the "jump" occur at finite frequency, the theoretical analysis predicts that the lowest out-of-plane mode ($\omega_{out}$) is always lower than its in-plane counterpart ($\omega_{in}$, see Fig. 3c,d), while the torsional vibrational modes ($\omega_\varphi$) are several orders of magnitude higher ($\omega_\varphi/\omega_{out} \propto L/r$, where $r$ and $L$ are the tube radius and length, respectively). Therefore, the mode that reaches zero frequency and dictates the ST transition is actually the out-of-plane mode. Accordingly, the lowest in-plane mode, which is usually the only mode we can observe in our measurements, will automatically exhibit a "jump" in the resonance frequency even before reaching zero (device VII, Fig. 3b and 3d). We shall also note that without twist, a negative jump (as in Fig. 1d left) cannot be accounted for. Occasionally, due to the rotational motion near the inflection point (representing the transition from up to down), we are able to detect evidence for the out-of-plane resonance frequency (marked by the arrows in Fig. 3a). Examining the CNT shape development (Fig. 3e-f) reveals how the in-plane and out-of-plane motions are strongly intertwined.

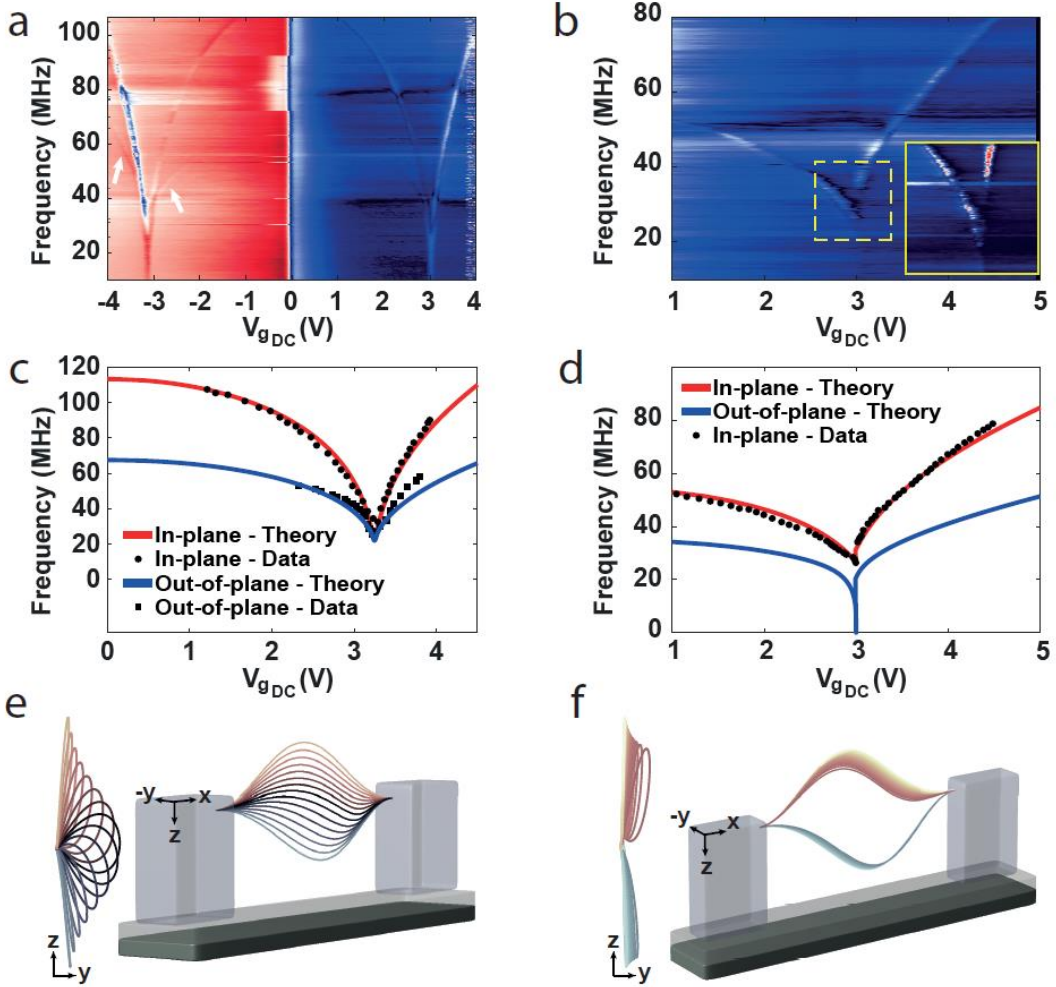

**Fig. 3: Hybrid in-plane and out-of-plane motion.** *(a) Resonance frequency measurement of device VII, revealing both the out-of-plane (lower, marked by the arrows) and in-plane (second) modes where $b_c \lesssim b_0 < 0$. (b) Resonance frequency measurement of device VIII, in which $b_0 < b_c < 0$. Only the lowest in-plane mode is detected. Inset is a close-up on the "jump" area marked by the dashed line. (c,d) Modelling the dynamic response of the same devices as in (a) and (b), respectively. The solid lines are the theoretical fitting to the experimental data (black dots\squares). (e) and (f) are the CNT static movement when increasing the gate voltage corresponding to the theoretical fitting in (c) and (d), respectively.*

**Broadening analysis**

Lastly, with the new experimental data at hand and having a model that reproduces the physical motion of the CNT, we address a long-standing question in the field of CNT resonators. The energy dissipation mechanisms of micro and nano resonators are an important characteristic of the device, commonly represented by their quality factor, Q, a quantity indicating how much energy dissipates in one period of oscillation. Unfortunately, the quality factor of CNT resonators at room temperature (~100) is much lower than anticipated (>1000) when considering known dissipation mechanisms[20,22,23,32–34].

An important experimental study showed that spectral broadening due to geometrical symmetry breaking can account for the low Q values at room temperature[23]. However, theoretical simulations suggest another possible mechanism ("fluctuation broadening"), in which thermal fluctuations should cause strong coupling between the in-plane and out-of-plane modes of motion, resulting from the dynamic built-in strain along the CNT[20]. This effect has never been proved experimentally due to the problematic estimation of the built-in strain in standard CNT resonators.

To study the dissipation mechanisms of the system, we calculate the quality factor according to $Q=f_0/\Delta f$, where $\Delta f$ is the full width at half maximum (FWHM) and $f_0$ is the resonance frequency (example of device IX in Fig. 4a). We compare the $\Delta f$ extracted from the data with a theoretical estimation of $\Delta f$ based on the fluctuation broadening theory (Fig. 4b, see supplementary text for details). In the supplementary text we also examine the broadening due to symmetry breaking[23,30], which has been previously shown to explain the low $Q$ at room temperature. We show that symmetry breaking cannot account by itself for the spectral broadening evidenced in our measurements (supplementary text).

The broadening due to the in-plane and out-of-plane coupling is estimated according to equation S33 in Ref. [20]:

$$\sigma_f = \frac{1}{2\pi}\left|\frac{\partial \omega_{ip}}{\partial x_{op}^2}\right| \cdot \sigma_{x_{op}^2} \qquad \text{(Eq. 1)}$$

where $x_{op}$ is the out-of-plane component of the CNT movement with thermal fluctuations given by $\sigma_{x_{op}^2} = \frac{k_B T}{m\omega_{op}^2}$, in which $\omega_{op}$ represents the lowest out-of-plane resonance mode.

The peak width is then estimated according to $\Delta f \sim 0.65\sigma_f$.

As evidenced in Fig. 4b, the fluctuation broadening theory yields remarkable compatibility to the experimental data. This is the most significant experimental evidence to date supporting this theory as the primary cause for low $Q$ in CNT resonators at room temperatures. Moreover, we can show that the other modes' fluctuations contribution to the broadening is negligible and that the lowest out-of-plane mode dominates (see supplementary information). To further examine this theory, we present a similar analysis on device IV at varying temperatures in the supplementary information (Fig. S9), also showcasing excellent agreement between the theoretical prediction and the experimental data.

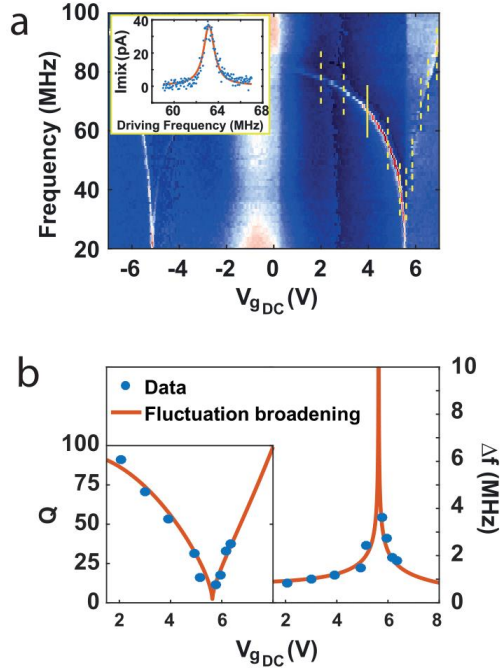

**Fig. 4: Spectral broadening at room temperature.** *(a) Resonance frequency measurement of device IX in the third category, where $b_c \lesssim b_0 < 0$. The yellow lines represent the frequency cross sections from which the data in (b) were extracted. Inset is an example taken at the yellow solid line. Blue dots are the data, orange solid line is a Lorentzian shape fit. (b) Quality factor (left) and spectral broadening (FWHM, right) extracted from (a) vs the static gate load (blue dots). The orange solid line is the theoretical calculation based on Eq. 1. The excellent fit implies that fluctuation broadening is the most significant mechanism to cause the low Q evidenced at room temperature.*

**Conclusions**

To conclude, thanks to a high quality fabrication process we were able to identify that there exist several types of carbon nanotube based resonators and that the main differentiating attribute is the shape of the as-fabricated buckled beam. We found that, unlike previous common-knowledge, it is possible to fabricate a CNT based resonator exhibiting Euler-Bernoulli snap-through bi-stability. The technological implication of the new classification is that while type A CNT resonators (Fig 2a) are the standard, the highest electrical frequency tunability is reached in type C (Fig. 2c). In addition, type D devices present a new double-well potential CNT resonator, where in both of the stable states the CNT is fully suspended with no physical contact to any surface. We speculate that this would open the possibility for fabricating integrated mechanical-resonator circuits for advanced computing as well as fast, durable and energy-efficient memory elements. On the theory side, we found that the existence of the snap-through bi-stability is associated with all three degrees of freedom of the CNT. Having a good experiment-model agreement, we also addressed the issue of anomalously low quality factor at room temperature and showed that nonlinear mode coupling accounts well for the spectral broadening in our devices, in agreement with the fluctuation broadening theory.

**Methods**

Fabrication

The geometric structure of our devices is presented in Fig. 1a. As a substrate, we used a silicon wafer with a layer of 500 nm silicon oxide on top. All patterning stages were done by standard photolithography or electron beam lithography procedures. We evaporated Cr/Pt 5nm/35nm for the source and drain (SD) electrodes, after which we etched the $SiO_2$ using buffer oxide etchant (BOE) to create the local-gate trench (see supplementary text for details). Then, another step of Cr/Pt 5nm/35nm evaporation was performed to create the self-aligned local gate. Finally, we deposited patterned ferritin catalyst near the SD electrodes for the CNT growth. The CNTs were grown by chemical vapor deposition (CVD) under a constant flow of 0.5/0.5 SLPM $H_2$/$CH_4$ at 900°C for 20min, utilizing the fast heating growth technique[35].

Resonance Frequency Measurements

The resonance frequency measurements were conducted using the mixing technique[22], presented schematically in Fig. 1a. The actuation of the CNT is done by applying an electrostatic force from the local gate:

$$F = \frac{1}{2}\frac{\partial C_g}{\partial z}V_g^2 = \frac{1}{2}\frac{\partial C_g}{\partial z}\left(V_g^{DC} + V_g^{AC}\cos(\omega t)\right)^2 \quad \text{(Eq. 3)}$$

where $C_g(z)$ is the capacitance between the CNT and the local-gate electrode, $V_g$ is the local-gate voltage, which is a combination of a DC voltage $V_g^{DC}$ superimposed with a driving voltage $V_g^{AC}$ excited at frequency $\omega$.

For the mixing, we apply a second driving voltage between the SD electrodes, $V_{sd}$:

$$V_{sd} = V_{sd}^{AC} \cos(\omega t + \Delta\omega t) \quad \text{(Eq. 4)}$$

where $V_{sd}^{AC}$ is the driving voltage, and $\Delta\omega$ is the offset frequency (in our experiments, few kHz). Utilizing the fact that the CNT serves as a mixer, we can actuate it at $\omega$ and detect its resonance frequencies through the current at the offset frequency $\Delta\omega$:

$$I_{mx}^{\Delta\omega} = \frac{1}{2} \frac{\partial G^{DC}}{\partial V_g^{DC}} \left( \frac{1}{C_g} \frac{\partial C_g}{\partial z} V_g^{DC} \delta z + V_g^{AC} \right) V_{sd}^{AC} \quad \text{(Eq. 5)}$$

where $G^{DC}$ is the DC conductance of the CNT and $\delta z$ is the time dependent vibrational motion of the CNT.

**Data availability**

The data that support the findings of this study are available from the corresponding author upon reasonable request.

**Acknowledgements**. This study was supported by the ISF (Grant No. 1854/19), and the Russell Berrie Nanotechnology Institute. The work made use of the Micro Nano Fabrication Unit at the Technion. S.R. acknowledges support by the Council for Higher Education and the Russel Berrie scholarships.

**Author contributions**. S.R. and T.T. have equal contribution to this work. S.R., T.T., and Y.E.Y. designed the experiment. S.R. and T.T. took part in device fabrication, performed the experiments and analysis. M.S. developed the fabrication process. S.S.

assisted in the theoretical modelling and performed finite element simulations. S.R., T.T., and Y.E.Y. wrote the manuscript, with input from all authors.

# Mode coupling, bi-stability, and spectral broadening in buckled nanotube resonators

# Supplementary information

S. Rechnitz[*,†], T. Tabachnik[*,†], M. Shlafman[†], S. Shlafman[†], and Y. E. Yaish[†]

[†]Andrew and Erna Viterbi Faculty of Electrical Engineering, Technion, Haifa 32000, Israel.

*These authors contributed equally to this work.

**Supplementary text contents:**

1. Fabrication and control of buckled CNT resonators
2. Influence of initial buckling
3. Snap-through buckling evidence in conductance measurements
4. Theoretical modelling
5. Calculation of the resonance frequencies and spectral broadening
6. Fluctural broadening calculation at varying temperatures
7. Nonlinear oscillator equation and symmetry breaking

1. <u>Fabrication and control of buckled CNT resonators</u>

   The fabrication process is detailed in the Methods section. The BOE etching is isotropic. Therefore, in addition to creating the trench between the source and drain electrodes, it also creates a small undercut underneath them. Then, another step of evaporation is performed, such that the final SD electrodes consist of a Cr/Pt/Cr/Pt 5nm/35nm/5nm/35nm heterostructure, with suspended edges. At the final stage of the CVD growth, the sample is heated to a temperature of 900°C, and thus the suspended metallic parts of the heterostructure will deform at the edges, due to thermal expansion coefficient mismatch. We believe that this deformation causes the CNT to favor upward buckling during the growth. To support our hypothesis, we examined the BOE etch depth vs the device classification (Fig. S1) and the device classification vs SD deformation (Fig. S2). Although the yield is not 100%, both graphs together with Fig. S3 demonstrate that we can control the initial buckling relatively well by the BOE etch step of the fabrication.

   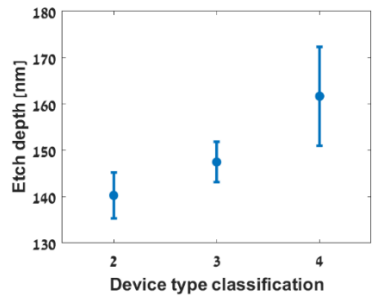

   **Fig. S1.** *The effect of the BOE etching depth (depth measured via AFM, reflecting on the etch time) on the classification of the device, for upward buckled CNT resonators. Data taken from all fabricated devices, most of which are not presented in the article.*

   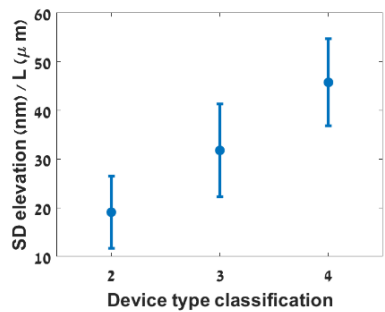

**Fig. S2.** *The effect of Source and Drain deformation (measured in AFM, normalized by the device length) on the classification of the device for upward buckled CNT resonators.*

2. Influence of initial buckling

    As explained in the main text (Fig. 2), the initial configuration of the CNT determines which type of behavior is to be expected. To support this claim, Fig. S3 presents the fitted $b_0$ vs classification for the devices in Fig. 2. One can clearly observe that as $b_0$ decreases (absolute value increases), the classification changes according to our qualitative explanation in the main text (Fig. 2).

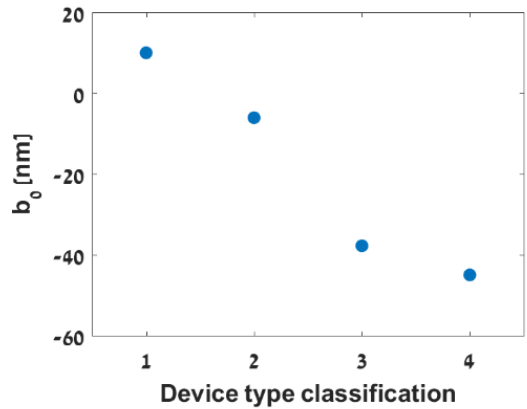

**Fig. S3.** *Initial midpoint elevation extracted from the fitting for the devices presented in Fig. 2, divided by their classification.*

Furthermore, to establish that the extracted initial buckling from the fitted data agrees with the real initial CNT displacement, we present a comparison of the theoretical $b_0$ with the SEM images of two devices in the third category (Fig. S4). From the SEM image in Fig. S4a we extracted initial midpoint elevation of $b_0=47\pm2$nm, and from the theoretical fitting (Fig. S4c) we extracted initial midpoint elevation of $b_0=42$nm. From the SEM image in Fig. S4d we extracted initial midpoint elevation of $b_0=43\pm4$nm, and from the theoretical fitting (Fig. S4f) we extracted initial midpoint elevation of $b_0=37.8$nm. We shall point out that the typical buckling heights are less than 5% of the tube length and are hence hard to notice in the SEM images.

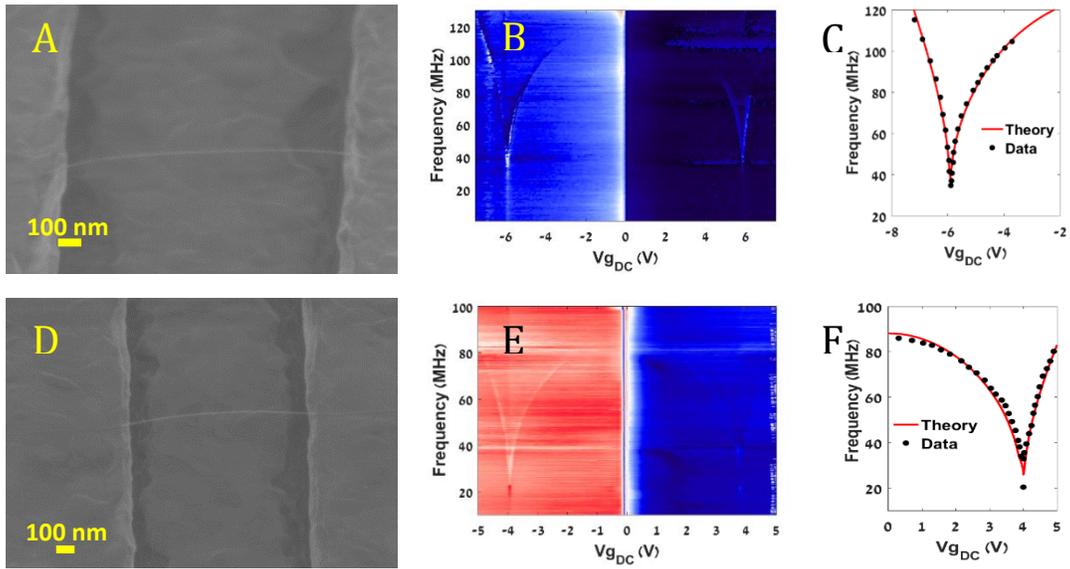

**Fig. S4.** *(a,d) SEM images of initially buckled upward CNTs (devices X and V, respectively), retrieved at a 75º and 70º (respectively) angle from the perpendicular to the surface. (b,e) Resonance frequency measurements of the same devices as in (a) and (d), respectively, typical of the third category (obtained prior to the SEM measurements). Measurement (e) is the same as Fig. 2c. (c,f) Theoretical fitting (red line) to the data (black dots) extracted from the resonance measurement in (b,e), respectively.*

3. <u>Snap-Through Buckling Evidence in Conductance Measurements</u>

The physical configuration of the CNT directly affects the capacitance between the CNT and the local-gate. Therefore, the snap-through (ST) and release transitions should be accompanied by capacitance modification. This change in the capacitance will modulate the charge on the CNT, translating into a change in the CNT conductance. Hence, at the unstable ST point, the abrupt mechanical transition causes a noticeable discontinuity in the conductance measurements, characterized also by hysteresis. Fig. S5 presents the transfer characteristics curve of device II. Along with the common small band gap behavior, we observe additional features of discontinuities and hysteresis (marked by the dashed lines), corresponding to the resonance "jumps" in Fig. 1d.

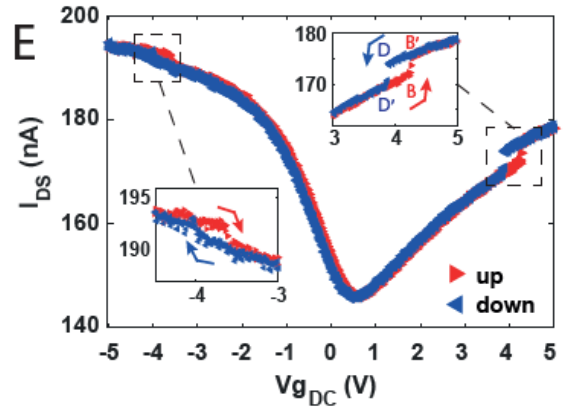

**Fig. S5.** *Conductance measurement of device II, exhibiting the snap-through and release phenomena. Insets: Zoom-in on areas marked by dashed rectangles.*

4. Theoretical Modelling

To analyze the static response of the system, we begin from the following relations of the force moment, **M**[1]:

$$\mathbf{M} = EI\mathbf{t} \times \frac{d\mathbf{t}}{dl} + \mathbf{t}GJ\tau$$

$$\frac{d\mathbf{M}}{dl} = \mathbf{F} \times \mathbf{t} \qquad \text{S1}$$

where $E$ is the CNT Young's modulus, $I$ is the moment of inertia, $\mathbf{t}$ is the unit vector tangential to the CNT ($\mathbf{t} = \frac{d\mathbf{r}}{dl}$), $G$ is the shear modulus, $J$ is the polar moment of inertia, and $\tau = \mathbf{\Omega} \cdot \mathbf{t}$ where $\mathbf{\Omega}$ is the moment of the forces acting on a cross section of the tube, thus $\tau = \frac{d\varphi}{dl}$ is the twisting component in the tangential direction and $\varphi(l)$ is the twist angle along the tube.

Taking the second derivative $d^2\mathbf{M}/dl^2$ results in:

$$EI\frac{d\mathbf{t}}{dl} \times \frac{d^2\mathbf{t}}{dl^2} + EI\mathbf{t} \times \frac{d^3\mathbf{t}}{dl^3} + GJ\tau\frac{d^2\mathbf{t}}{dl^2} + 2GJ\frac{d\tau}{dl}\frac{d\mathbf{t}}{dl} + GJ\frac{d^2\tau}{dl^2}\mathbf{t} = \frac{d\mathbf{F}}{dl} \times \mathbf{t} + \mathbf{F} \times \frac{d\mathbf{t}}{dl}$$

S2

Substituting $\frac{d\mathbf{F}}{dl} = -\vec{\kappa}$, where $\vec{\kappa} = \kappa_z$ the external force (acting in the z direction only), assuming $dl \simeq dx$ (shallow arch) and separating into x,y,z components yields the following set of equations which describe the CNT static motion:

$$\hat{x}: EI\left(y'z'''' - z'y'''' + y''z''' - z''y'''\right) + GJ\varphi''' = y'\kappa_z - F_z y''$$

$$\hat{y}: EIz'''' - GJ\left(\varphi'y''' + 2\varphi''y'' + \varphi'''y'\right) = \kappa_z - Pz'' + \frac{EA}{2L}z''\int_0^L \left(z'^2 + y'^2 + r^2\varphi'^2\right)dx$$

$$\hat{z}: EIy'''' + GJ\left(\varphi'z''' + 2\varphi''z'' + \varphi'''z'\right) = -Py'' + \frac{EA}{2L}y''\int_0^L \left(z'^2 + y'^2 + r^2\varphi'^2\right)dx$$

S3

where we substituted $F_x = -P + \frac{EA}{2L}\int_0^L \left(z'^2 + y'^2 + r^2\tau^2\right)dx$, the tension along the tube, in which $P$ is the initial axial tension at $V_g^{DC} = 0$. $z(x,t), y(x,t)$ are the in-plane and out-of-plane deflections along the beam, respectively, and $\varphi(x,t)$ is the twist angle along the tube. $A$ is the CNT cross-section area, $L$ is the horizontal distance between the clamped ends of the suspended CNT, $r$ is the CNT radius and $\kappa_z$ is the electrostatic force exerted by the local-gate.

For the dynamic response analysis, we add the acceleration term to the $\hat{y}, \hat{z}$ equations to receive the Euler-Bernoulli beam equations. For the torsional vibrations we use the first derivative $\frac{d\vec{M}}{dl} - \vec{F}\times\vec{t} = \rho I_x \frac{\partial^2 \varphi}{\partial t^2}$. Finally, we can describe the dynamic motion of the CNT by the following set of equations:

$$\hat{x}: EIy'z''' - EIz'y''' + GJ\varphi'' + F_z y' - \rho I_x \ddot{\varphi} = 0$$
$$\hat{y}: EIz'''' - GJ\left(\varphi'y''' + 2\varphi''y'' + \varphi'''y'\right) + \rho A\ddot{z} - \kappa_z + Tz'' = 0 \quad\quad S4$$
$$\hat{z}: EIy'''' + GJ\left(\varphi'z''' + 2\varphi''z'' + \varphi'''z'\right) + \rho A\ddot{y} + Ty'' = 0$$

where $t$ is time, $I_x$ is the torsional moment of inertia, $\rho$ is the mass density, and $T$ is the tension, given by $T = -P + \frac{EA}{2L}\int_0^L \left(z'^2 + y'^2 + r^2\varphi'^2\right)dx$, as before.

We define the tube's deflection as a superposition of the initial buckling (subscript 0), the static deflection due to the DC voltage (subscript s) and the dynamic oscillation due to the AC actuation (subscript d), as illustrated for the in-plane component in Fig. S6:

$$z(x,t) = z_0(x) + z_s(x) + z_d(x,t)$$
$$y(x,t) = y_0(x) + y_s(x) + y_d(x,t)$$
$$\varphi(x,t) = \varphi_0(x) + \varphi_s(x) + \varphi_d(x,t)$$

Doubly clamped boundary conditions were imposed.

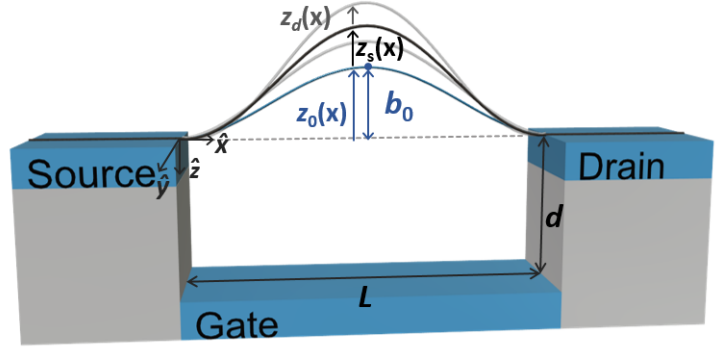

**Fig. S6.** *Schematic illustration for a doubly clamped suspended CNT with in-plane movement, represented by the superposition of the initial buckling ($z_0$), static deflection ($z_s$) and dynamic deflection ($z_d$).*

Eqs. S3 and S4 are nonlinear integro-differential equations, and therefore finding an exact analytical solution is difficult. However, by discretizing the system into finite degrees of freedom using a reduced order model (ROM) instead of solving a continuous system, an approximated solution can be found. For this task, we chose the Galerkin method[2], in which the beam deflection is approximated as a linear combination of eigenmodes of the linear Euler-Bernoulli equation for a doubly clamped straight beam, noted by $\xi_i(x)$:

$$z(x,t) = \sum_{i=1}^{n} q_i(t)\xi_i(x) = \sum_{i=1}^{n}\left[q_{0,i} + q_{s,i} + q_{d,i}(t)\right]\xi_i(x)$$

$$y(x,t) = \sum_{i=1}^{m} v_i(t)\xi_i(x) = \sum_{i=1}^{n}\left[v_{0,i} + v_{s,i} + v_{d,i}(t)\right]\xi_i(x)$$

$$\varphi(x,t) = \sum_{i=1}^{m} \varphi_i(t)\xi_i(x) = \sum_{i=1}^{n}\left[\varphi_{0,i} + \varphi_{s,i} + \varphi_{d,i}(t)\right]\xi_i(x)$$

The discretization is therefore realized by taking a finite number of eigenmodes. Specifically, by comparing with a finite element method solution, we find that only the first two modes suffice:

$$\begin{aligned}
z_0(x) &= q_{01}\xi_1(x) + q_{02}\xi_2(x) & y_0(x) &= v_{01}\xi_1(x) + v_{02}\xi_2(x) & \varphi_0(x) &= \varphi_0 x + \varphi_{01}\xi_1(x) + \varphi_{02}\xi_2(x) \\
z_s(x) &= q_{1s}\xi_1(x) + q_{2s}\xi_2(x) & y_s(x) &= v_{1s}\xi_1(x) + v_{2s}\xi_2(x) & \varphi_s(x) &= \varphi_{1s}\xi_1(x) + \varphi_{2s}\xi_2(x) \\
z_d(x,t) &= q_{1d}(t)\xi_1(x) + q_{2d}(t)\xi_2(x) & y_d(x,t) &= v_{1d}(t)\xi_1(x) + v_{2d}(t)\xi_2(x) & \varphi_d(x,t) &= \varphi_{1d}(t)\xi_1(x) + \varphi_{2d}(t)\xi_2(x)
\end{aligned}$$

This process transforms Eqs. S3 and S4 into a set of algebraic equations, which are solved numerically. We solve Eqs. S3 for the initial buckling and static motion ($q_0, q_s, v_0, v_s, \varphi_0, \varphi_s$) and only then solve Eqs. S4 for the resonance frequencies (i.e., the eigenvalues of the homogeneous system of equations) at the specific static position.

We shall emphasize that the CNT initial configuration is not restricted to be solely in-plane. We allow initial out-of-plane deflection by taking $y_0 \neq 0$. Nonetheless, our SEM images and theoretical fitting reveal that the out-of-plane initial buckling is much smaller compared to the in-plane buckling ($y_0 \ll z_0$). Hence, the qualitative differentiation between device types based solely on the in-plane $b_0$ in Fig. 2 is sufficient. Quantitatively, in order to achieve the good fitting, we must include non-zero out-of-plane deflection, even if it is small.

Fig. S7 presents a SEM image of the same device as Fig. S4a, taken perpendicular to the surface (from above), in which the CNT appears nearly straight, meaning that the out-of-plane component of the initial beam shape ($y_0$) is very small compared to the in-plane component ($z_0$) apparent in Fig. S4a.

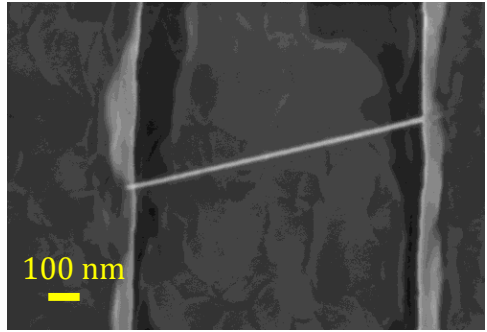

**Fig. S7.** *Top view SEM image of the same CNT as in Fig. S4a, retrieved perpendicular to the sample surface. The beam appears straight, implying that the out-of-plane component of the initial beam shape ($y_0$) is much smaller than the in-plane curvature ($z_0$) apparent in Fig. S4a.*

5. Calculation of the resonance frequencies and spectral broadening

As explained in the main text, the built-in strain along the nanotube affects its resonance frequencies. Since the resonance frequencies of the out-of-plane modes are always lower than the in-plane modes, we expect that the thermal fluctuations of the out-of-plane modes $\sigma_{x_{op}^2} = \frac{k_B T}{m\omega_{op}^2}$ will affect both the in-plane resonance frequencies, $f_{ip}$, and their spectral broadening, $\Delta f_{ip}$ [3]. Throughout the mode coupling discussion, we shall neglect the coupling to the torsional modes since, as explained in the main text, their resonance frequencies are much higher and therefore their effect on the in-plane modes is negligible.

It is customary to transform equations S3 and S4 to be dimensionless. We normalize the equations according to $x \to \frac{x}{L}$, $y \to \frac{y}{d}$, $z \to \frac{z}{d}$, $r \to \frac{r}{d}$, $t \to \frac{t}{t_0}$, where $d$ is the trench depth and $t_0 \triangleq \sqrt{\frac{\rho A L^4}{EI}}$, so the out-of-plane motion is represented by the dimensionless parameter $v_d = \frac{y}{d}$. The dimensionless algebraic in-plain equations include coupling between the in-plane and out-of-plane modes in the form of $\propto q_d \left( v_{1d}^2 \int \xi_1'^2 + v_{2d}^2 \int \xi_2'^2 + ... \right)$, expressing the built-in strain resulting from the vibrations of the out-of-plane modes.

Barnard *et al.* show in Ref. 3 that for a buckled beam, the in-plane resonance fluctuations, $\sigma_f$, are related to the thermal fluctuations of the out-of-plane mode, $\sigma_{y^2} = \frac{k_B T}{m\omega_{op}^2}$, according to:

$$\sigma_f = \frac{1}{2\pi} \left| \frac{\partial \omega_{ip}}{\partial y^2} \right| \cdot \sigma_{y^2} \qquad \text{S5}$$

where $k_B$ is the Boltzmann constant, $T$ is temperature and $m\omega_{op}^2$ is the effective spring constant of the out-of-plane mode.

Our theoretical model allows us to calculate both the out-of-plane resonance frequencies, and the derivative $\left|\frac{\partial \omega_{ip}}{\partial v_d^2}\right|$ for every static load, according to

$$\left|\frac{\partial \omega_{ip}}{\partial v_d^2}\right| = \frac{1}{2\omega_{ip}}\left(\frac{\partial \omega_{ip}^2}{\partial v_d^2}\right).$$

Substituting these results into Eq. S5 and using the relation $\Delta f_{FWHM} \sim 0.65 \sigma_f$ yields the excellent fit for device IX in Fig. 4. Another example of a good fit for a second CNT (device I, Fig. 1c) is presented in Fig. S11.

One should wonder how dominant is the lowest out-of-plane mode compared to higher out-of-plane modes as well as higher in-plane modes[4]. To take into account an additional out-of-plane mode, we simply sum their contributions according to:

$$\sigma_{y^2} = \sqrt{\left(\frac{k_B T}{m\omega_{op1}^2}\right)^2 + \left(\frac{k_B T}{m\omega_{op2}^2}\right)^2} \qquad \text{S6}$$

It turns out that the contribution of the second mode is negligible (Fig. S8). This result is not surprising since the lowest out-of-plane resonance frequency is very close to the lowest in-plane mode, whereas the second out-of-plane and second in-plane resonance frequencies are significantly higher. For example, at $V_g=0$, $\omega^2_{op2}/\omega^2_{op1}=17.52$ and $\omega^2_{ip2}/\omega^2_{op1}=18.12$, and at the inflection point ($V_g=5.532V$) $\omega^2_{op2}/\omega^2_{op1}=25.55$ and $\omega^2_{ip2}/\omega^2_{op1}=26.76$.

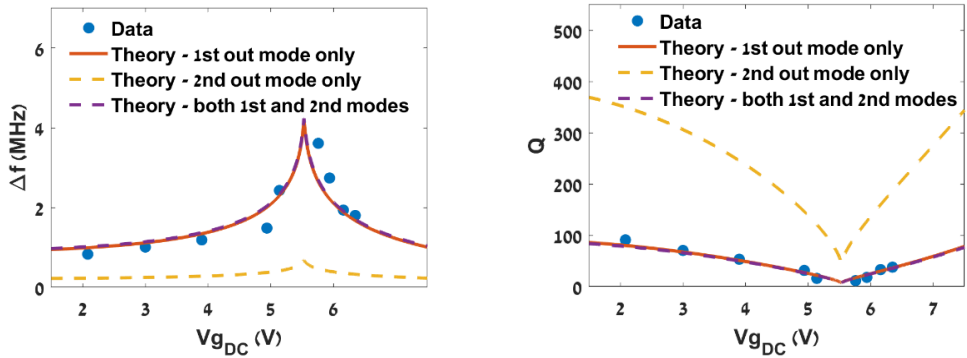

**Fig. S8.** *Spectral broadening (FWHM, left) and quality factor (right) extracted from Fig. 4a vs the static gate load (blue dots). The orange solid line is the theoretical calculation based on Eq. 1, considering only the 1st out-of-plane mode. The dashed yellow line is the theoretical calculation, considering only the 2nd*

*out-of-plane mode. And the dashed purple line is the theoretical calculation summing the contribution of both the 1$^{st}$ and 2$^{nd}$ out-of-plane modes. The excellent fit implies that fluctuation broadening is the most significant dissipation mechanism in the system, and the similarity between the orange and purple lines implies that the coupling to the first mode governs the broadening.*

6. Fluctural broadening calculation at varying temperatures

Since fluctuation broadening is based on the equipartition theorem, the broadening depends on temperature. Hence, to further investigate this theory, we decided to repeat the same measurements and spectral broadening analysis at varying temperatures. At each temperature, we performed a resonance frequency measurement vs gate voltage (example obtained at T=38°K is presented in Fig. S9a). From the theoretical fitting to this data (Fig. S9b), we calculated the strain gate voltage dependence at the different temperatures. Then we used the same methodology for calculating the theoretical broadening due to the first out-of-plane mode using Eq. 1. The experimental spectral broadening was extracted from a general Lorentzian fit to the resonance peak measured at each temperature (Fig. S9a inset). The comparison between the broadening extracted directly from the measurement and the broadening estimated according to the theory at the same gate voltage and temperature is presented in Fig. S9c. The compatibility between experiment and theory is quite remarkable.

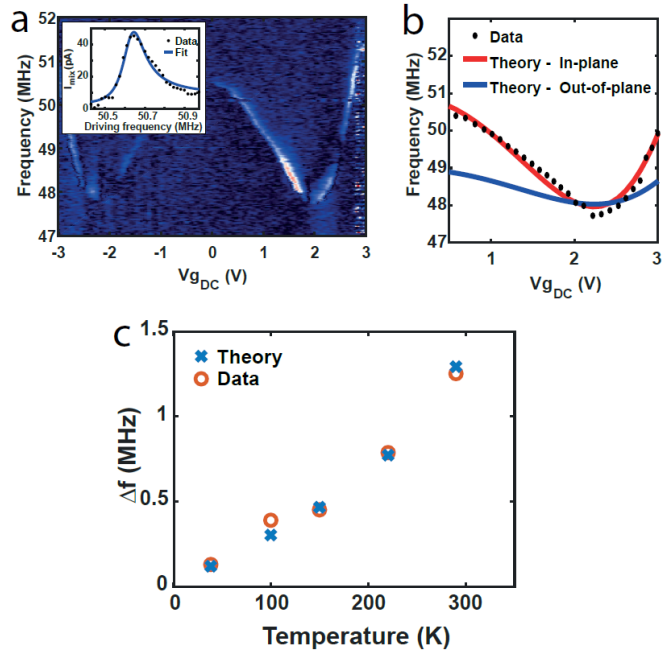

**Fig. S9.** *(a) Resonance frequency measurement of device IV in the second category, obtained at T=38°K. Inset is a frequency cross section obtained at $V_{gDC}$=0.8V, from which the left-most data mark in (c) was extracted. Black dots are the data, blue solid line is a Lorentzian shape fit. (b) Modelling the dynamic response of the measurement in (a), from which the left-most theoretical mark in (c) was calculated. The solid lines are the theoretical fitting to the experimental data (black dots). (c) Comparison between the experimental spectral broadening (FWHM) extracted from resonance peaks measured at varying*

*temperatures from device IV (orange marks) vs. the theoretical prediction according to Eq. 1 (blue marks).*

7. <u>Nonlinear oscillator equation and symmetry breaking</u>

For this discussion, we shall take only the first in-plane mode and neglect the out-of-plane and torsion coupling terms. In this case, if we add the damping term, $\gamma \frac{\partial z_d}{\partial t}$, the dynamic beam equation for the in-plane modes is:

$$EI\frac{\partial^4 z_d}{\partial x^4} + \rho A \frac{\partial^2 z_d}{\partial t^2} + \gamma \frac{\partial z_d}{\partial t} - \left(-P + \frac{EA}{2L}\int_0^L \left(\frac{\partial z_0}{\partial x} + \frac{\partial z_s}{\partial x}\right)^2 dx\right)\frac{\partial^2 z_d}{\partial x^2} - \left(\frac{EA}{2L}\int_0^L \left[2z_0'z_d' + 2z_s'z_d' + z_d'^2\right]dx\right)\frac{\partial^2 z}{\partial x^2} = \kappa_d$$

where $\kappa_d$ is the dynamic force exerted by the local-gate.

Substituting $z_0(x) = q_{01}\xi_1(x)$, $z_s(x) = q_{1s}\xi_1(x)$, $z_d(x,t) = q_{1d}(t)\xi_1(x)$, multiplying the equation by $\xi_1(x) \cdot L$ and integrating over the beam length results in:

$$\frac{EI}{L^2} q_{1d} \int \xi_1''\xi_1'' + \rho A L \frac{d^2 q_{1d}}{dt^2} \int \xi_1^2 + cL\frac{dq_{1d}}{dt}\int \xi_1^2 - \left(-P + \frac{EA}{2L^2}(q_{01}+q_{1s})^2\int \xi_1'\xi_1'\right)q_{1d}\int \xi_1'\xi_1'$$

$$+\frac{EA}{2L^2}\left(\int \xi_1'\xi_1'\right)^2 (q_{01}+q_{1s}+q_{1d})\left(2(q_{01}+q_{1s})q_{1d}+q_{1d}^2\right) = \kappa_{zd}\int \xi_1$$

S8

Rearranging the terms in Eq. S8 to the powers of $q_{1d}(t)$ and its derivatives will lead to the following equation:

$$\rho A L \int \xi_1^2 \cdot \frac{d^2 q_{1d}}{dt^2} + cL\int \xi_1^2 \cdot \frac{dq_{1d}}{dt} + \left[\frac{EI}{L^2}\int \xi_1''\xi_1'' - P\int \xi_1'\xi_1' + \frac{3EA}{2L^2}\left(\int \xi_1'\xi_1'\right)^2(q_{01}+q_{1s})^2\right]q_{1d}$$

$$+\frac{3EA}{2L^2}\left(\int \xi_1'\xi_1'\right)^2(q_{01}+q_{1s})q_{1d}^2 + \frac{EA}{2L^2}\left(\int \xi_1'\xi_1'\right)^2 q_{1d}^3 = \kappa_{zd}\int \xi_1$$

S9

By defining $u(t) \equiv q_{1d}(t)$ and the following coefficients:

$$m = \rho A L \int_0^L \xi_1^2 \, dx$$

$$\gamma = cL \int_0^L \xi_1^2 \, dx$$

$$k = \frac{EI}{L^2}\left(\int_0^L \left(\frac{d^2\xi}{dx^2}\right)^2 dx\right) - P\left(\int_0^L \left(\frac{d\xi}{dx}\right)^2 dx\right) + \frac{3EA}{2L^2}\left(\int_0^L \left(\frac{d\xi}{dx}\right)^2 dx\right)^2 (q_{01} + q_{1s})^2$$

$$k_2 = \frac{3EA}{2L^2}\left(\int_0^L \left(\frac{d\xi}{dx}\right)^2 dx\right)^2 (q_{01} + q_{1s})$$

$$k_3 = \frac{EA}{2L^2}\left(\int_0^L \left(\frac{d\xi}{dx}\right)^2 dx\right)^2$$

$$f = \kappa_d \int_0^L \xi(x)\, dx$$

$$\tag{S10}$$

Eq. S8 becomes:

$$m\frac{d^2 u}{dt^2} + \gamma \frac{du}{dt} + ku + k_2 u^2 + k_3 u^3 = f \tag{S11}$$

By examining carefully the coefficients of Eq. S11, it is noticeable that introducing initial curvature $q_{01}$ to the beam has double impact, as it shifts the value of the linear spring constant (i.e. $k$) and at the same time creates quadratic nonlinearity (i.e. $k_2$). This explains why the behavior of arch shaped bi-stable systems is dictated by the quadratic nonlinearity, which always results in softening of the natural frequency[5].

Experimentally, when the actuation power is small, the nonlinear terms in Eq. S11 are negligible and the linear response of the dynamic problem (Figs. 1-4) is mostly dictated by $k$, as described in the main text. Raising the actuation power and thus increasing the mechanical vibration at a given DC gate voltage, we observe the anticipated softening behavior as well as hysteresis, both before and after the ST transition (Fig. S10). This phenomenon is different from previous results which usually observe hardening at low static loads, and softening at higher gate voltages[6,7].

The solution of Eq. S10 is given by[2,8]:

$$\left[\frac{\gamma^2}{4}+\left((\omega_d-\omega_0)-\frac{3}{8}\frac{k_{eff}}{\omega_0}u^2\right)^2\right]u^2=\frac{f_d^2}{4\omega_0^2} \qquad \text{S12}$$

where $\omega_d$ and $f_d$ are the frequency and amplitude of the excitation $f$, respectively, and $k_{eff}=k_3-\frac{9k_2^2}{10\omega_0^2}$ is the effective nonlinear coefficient. We can extract $k_{eff}$ from the resonance shift vs. $u^2$ at a constant DC voltage, according to the relation[7]:

$$\omega_{max}-\omega_0=\frac{3}{8}\frac{k_{eff}u^2}{\omega_0} \qquad \text{S13}$$

where $\omega_{max}$ is the driving frequency for which we detect maximum current, corresponding to maximum deflection. The amplitude $u$ of the vibration is extracted from the maximum current in the frequency response, according to[6]: $I_{peak}=\frac{1}{2}\frac{\partial G}{\partial V_g}\cdot\delta v_{sd}\cdot V_{gDC}\cdot\frac{C_g'}{C_g}\cdot u$.

$k_2$ linearly depends on the CNT static midpoint displacement $(q_{01}+q_{1s})$. Therefore, near the ST point, when the deflection is small, the quadratic nonlinearity is negligible and $k_{eff}\approx k_3$. However, farther from the snap (either before or after), $k_2$ is expected to be dominant, resulting in softening behavior ($k_{eff}<0$) at both low and high static loads. For example, at $V_{gDC}$=4.8 V (Fig. S10d), we receive an effective nonlinear coefficient of $k_{eff}=(-8.1\pm0.1)\cdot10^{32}\ s^{-2}\cdot m^2<0$, in agreement with the observed softening behavior.

Using the $k_{eff}$ and $\gamma$ coefficients retrieved from the data, we calculated the anticipated response of the system and obtained excellent fit of Eq. S11 to the measured Duffing-like behavior (Figs. S10b and S10d).

This analysis also allows us to estimate the spectral broadening due to symmetry breaking, which has been proposed as an alternative mechanism responsible for the low Q of CNT resonators at room temperature[7]. This geometric symmetry breaking actually refers to the geometric quadratic nonlinearity, $k_2$. We calculate this effect according to equation S14 of Ref. 6:

$$\Delta f = \frac{3 k_{eff} k_B T}{8 \pi m \omega_0^3} \qquad \text{S14}$$

For example, substituting $\omega_0$ and $k_{eff}$ measured at $V_{gDC} = 4.8V$ into Eq. S14 results in $\Delta f \simeq 0.52\,MHz$ and $Q \simeq 201$, which is 28% smaller than the measured broadening of $\Delta f_{exp} = 1.86\,MHz$ and yielding Q nearly four times larger than $Q_{exp} = 56.13$. A similar result is achieved for all five gate voltages marked in Fig. S10a and is plotted in Fig. S11, implying that although characterized by substantial geometric nonlinearity, symmetry breaking is not the major broadening mechanism in our devices.

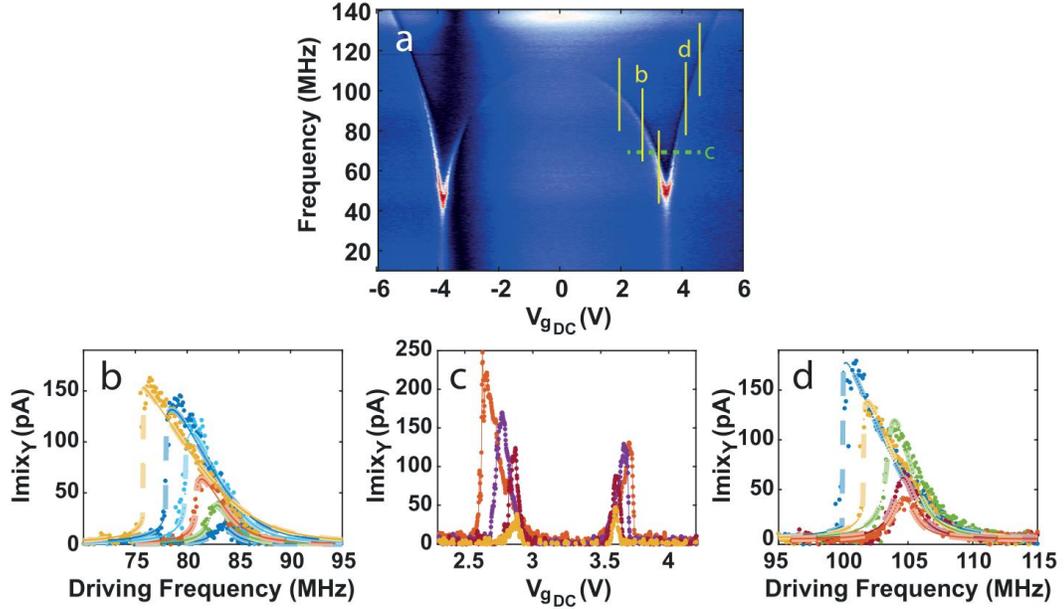

**Fig. S10. Duffing-type behavior.** *(a) Resonance frequency measurement of device I (same as Fig.1c) in the linear regime. (b) and (d) present examples of two out of the five frequency cross sections marked in (a) before (at $V_{gDC}$=2.5V) and after (at $V_{gDC}$=4.8V) the IP transition, respectively, for a series of increasing $V_{SD}$ excitations, revealing non-linear dynamic response of the system, governed by softening. Dots are the sampled data and lines are the theoretical fit*

*from solving equation S12. (c) Gate voltage cross sections in low and high excitations (marked by the green line in (a)), also displaying softening behavior.*

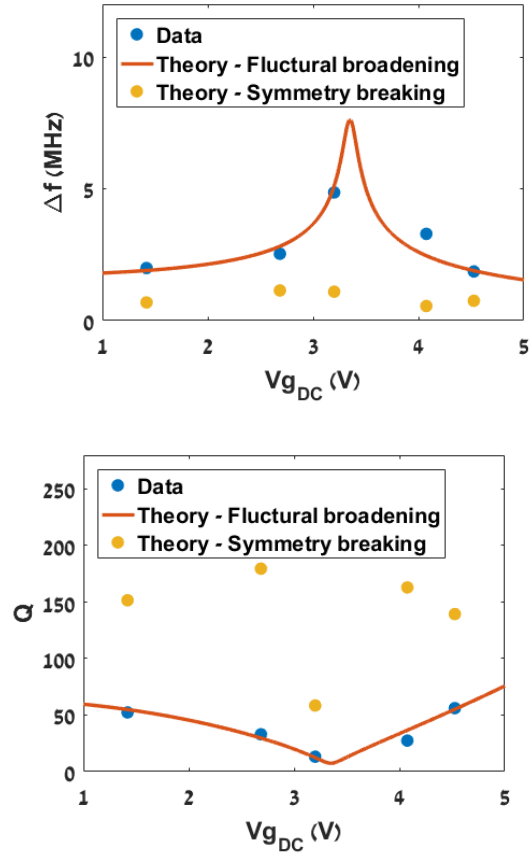

**Fig. S11.** *Peak width (left) and quality factor (right) calculated for device I according to the fluctural broadening theory (Eq. 1, orange line) and according to the symmetry breaking theory (Eq. S14, yellow dots) vs. the experimental data (blue dots).*